\documentstyle[epsfig,aps,floats,preprint]{revtex}

\newcommand{\beq}{\begin{equation}}
\newcommand{\eeq}{\end{equation}}
\newcommand{\bea}{\begin{eqnarray}}
\newcommand{\eea}{\end{eqnarray}}

\def\laq{~\raise 0.4ex\hbox{$<$}\kern -0.8em\lower 0.62
ex\hbox{$\sim$}~}
\def\gaq{~\raise 0.4ex\hbox{$>$}\kern -0.7em\lower 0.62
ex\hbox{$\sim$}~}

\def \fb {\overline \phi}

\def \pfb {\Pi_{\fb}}
\def \pbe {\Pi_{\b}}

\def \ra {\rightarrow}
\def \la {\lambda}
\def \La {\Lambda}

\def \b {\beta}

\def \da {\delta}

\def \Om {\Omega}

\begin{document}
\par
\begingroup

\begin{flushright}
BA-TH/00-378\\
February 2000\\
gr-qc/0002080\\
\end{flushright}

\vspace{12mm}
{\large\bf\centering\ignorespaces
Birth of the Universe as anti-tunnelling\\
from the string perturbative vacuum
\vskip2.5pt}

\bigskip
{\dimen0=-\prevdepth \advance\dimen0 by23pt
\nointerlineskip \rm\centering
\vrule height\dimen0 width0pt\relax\ignorespaces
M. Gasperini
\par}

{\small\it\centering\ignorespaces
Dipartimento di Fisica, Universit\`a di Bari, \\
Via G. Amendola 173, 70126 Bari, Italy \\
and \\Istituto Nazionale di Fisica Nucleare, Sezione di Bari,
Bari, Italy \\
\par}

\par
\bgroup
\leftskip=0.10753\textwidth \rightskip\leftskip
\dimen0=-\prevdepth \advance\dimen0 by17.5pt \nointerlineskip
\small\vrule width 0pt height\dimen0 \relax

\begin{abstract}
The decay of the string perturbative vacuum, if triggered by a suitable,
duality-breaking dilaton potential, can efficiently proceed via the
parametric amplification of the Wheeler-De Witt wave function in
superspace, and can appropriately describe the birth of our Universe as
a quantum process of pair production from the vacuum.
\end{abstract}

\vspace{10mm}
\begin{center}
---------------------------------------------\\
\vspace {5 mm}
{\sl Essay written for the 2000 
Awards of the Gravity Research Foundation,}\\
{\sl and selected for Honorable Mention.}\\
To appear in {\bf Int. J. Mod. Phys. D}
\end{center}
\vspace{5mm}

\par\egroup
\thispagestyle{plain}
\endgroup



A consistent and quantitative description of the birth of our Universe is
one of the main goals of the quantum approach to cosmology. In the
context of the standard scenario, in particular, quantum effects are
expected to stimulate the birth of the Universe in a state approaching
the de Sitter geometric configuration, appropriate to inflation \cite{1}.
The initial cosmological state is unknown, however, and has to be fixed
through some ``ad-hoc" prescription. It follows that there are various
possible choices for the initial boundary conditions \cite{2,3,4},
leading in general to different quantum pictures of the early
cosmological evolution. 

In the context of the pre-big bang scenario \cite{5}, typical of string
cosmology, the initial state on the contrary is fixed, and has to
approach the string perturbative vacuum. The quantum decay of  this
initial state necessarily crosses the high-curvature, Planckian regime,
and can be appropriately described by a Wheeler-de Witt
(WDW) wave function \cite{6}, evolving in superspace. The birth of the
Universe may then be represented as a process of
scattering and reflection \cite{7,7a}, in an appropriate minisuperspace
parametrized by the metric and by the dilaton. In that case the pre-big
bang initial state -- emerging from the string perturbative vacuum -- 
simulates the boundary conditions prescribed for a process of
``tunnelling from nothing", in the context of the standard scenario
\cite{4}. It seems thus appropriate  to say that the above scattering
process describes the birth of the Universe as a {\em ``tunnelling from
the string perturbative vacuum"} \cite{7,8}. 

In a process of tunnelling, or quantum reflection, the WDW wave
function corresponding to our present cosmological configuration turns
out to be exponentially damped: the birth of the Universe from the 
string perturbative vacuum would thus appear to be a very unlikely 
(i.e., highly suppressed) quantum effect, according to the above
representation. In the string cosmology minisuperspace, however,
there are also other, more efficient ``channels"
of vacuum decay. The main  purpose of this paper is to show that, with
an appropriate model of dilaton potential,  the transition from the
pre-big bang to the post-big bang regime  may correspond to a
parametric amplification of the wave function, in such a way that
the birth of the Universe can be represented as a process of {\em
``anti-tunneling from the string perturbative vacuum"}. The name
``anti-tunnelling", which is synonymous of parametric amplification (a
well known effect in the theory of cosmological perturbations\cite{10})
follows from the fact that the transition probability in that case is
controlled by the inverse of the quantum-mechanical transmission
coefficient in superspace. 

In order to illustrate this possibility we shall use a quantum cosmology
model based on the lowest order, gravi-dilaton string effective action,
which in $d+1$ dimensions can be written as:  
\beq
S = -\frac{1}{2\,\lambda_s^{d-1}}\,\int\,d^{d+1}x\,\sqrt{|g|}\,e^{-\phi}
\,\left[R+(\nabla_\mu\phi)^2 +V (\phi,
g_{\mu\nu})\right]. 
\label{1}
\eeq
Here $\lambda_s$ is the
fundamental string length parameter, and $V$ is the (possibly non-local
and non-perturbative) dilaton potential.  Considering an isotropic,
spatially flat cosmological background,
\beq
\phi=\phi(t), ~~~~~~~~~~~~~
g_{\mu\nu} ={\rm diag} \left(g_{00}(t), -a^2(t) \da_{ij}\right), 
\label{2}
\eeq
with spatial sections of finite volume (a toroidal space, for instance),
it is convenient to introduce the variables
\beq
\b= \sqrt{d} \ln a, ~~~~~~~~~~~~~
\fb=\phi -\sqrt{d}\,\beta -\ln\,\int\,d^dx/\lambda_s^d, 
\label{3}
\eeq 
and the corresponding canonical momenta, defined in the cosmic
time gauge by: 
\beq
\Pi_{\beta}=\left({\da S\over \da\dot{\beta}}\right)_{g_{00}=1}=
\lambda_s\,\dot{\beta}\,e^{-\fb} , ~~~~~~~~~~~~
\Pi_{\fb}=\left({\da S\over \da\dot{\fb}}\right)_{g_{00}=1}=
-\lambda_s\,\dot{\fb}\,e^{-\fb} .
\label{4}
\eeq
The WDW equation, which implements the Hamiltonian constraint $H=
\da S/\da g_{00} =0$ in the two-dimensional minisuperspace spanned by
$\b$ and $\fb$, takes then the form \cite{7,7a,8}:
\beq
\left [ \partial^2_{\fb} - 
\partial^2_{ \beta}
+\lambda_s^2\,V(\b,\fb)\,e^{-2\fb}~ \right ]\, \Psi(\b,\fb)= 0 
\label{5}
\eeq
(we have used the differential representation $\Pi^2= -\nabla^2$). 

As is well known from low-energy, perturbative theorems, the
dilaton potential is strongly suppressed (with an istantonic law) in the
small coupling regime, so that the effective WDW potential appearing in
eq.(\ref{5})  goes to zero as we approach the flat, zero-coupling,
string perturbative vacuum, $\b \ra -\infty$, $\fb \ra -\infty$. In the
opposite regime of arbitrarily large coupling the dilaton potential is
unknown, but we shall assume in this paper that a possible growth of
$V$ is not strong enough to prevent the effective WDW potential from
going to zero  also at large positive
values of $\beta$ and $\fb$, so that $V \exp(-2\fb) \ra 0$ for $\b,\fb
\ra \pm \infty$. In this case, the asymptotic solutions of the WDW
equations (\ref{5}) can be factorized in the form of plane waves,
representing free energy and momentum eigenstates:
\beq
\Psi(\b,\fb) = \psi_k^{\pm}(\b)\psi_k^{\pm} ({\fb}) \sim e^{\pm ik \b
\pm ik \fb}, \label{6}
\eeq
where ($k>0$):
\beq
\pbe\psi_k^{\pm}(\b)=  \pm
k\psi_k^{\pm}(\b),~~~~~~~~~~~~~
\pfb\psi_k^{\pm}({\fb})= \pm 
k\psi_k^{\pm}({\fb})
\label{7}
\eeq
From a geometric point of view they represent, in minisuperspace, the
four branches of the classical, low-energy string cosmology solutions
\cite{5}, defined by the condition $\pbe = \pm \pfb$, and corresponding
to \cite{7,7a,8}:
\begin{itemize}
\item 
expansion, ~~~ $\pbe >0$, ~~~~~~~~~~~~~~~~~~
contraction,~~~ $\pbe <0$,  
\item
pre-big bang, ~~$\pfb <0$, ~~~~~~~~~~~~~~~  post-big 
bang,~~ $\pfb >0$. 
\end{itemize}

We now recall that, for an isotropic string cosmology solution \cite{5},
the dilaton is growing ($\dot \phi>0$) only if the metric is
expanding ($\dot\b >0$), see eq. (\ref{3}). If we impose, as our physical
boundary condition, that the Universe emerges from the string
perturbative vacuum (corresponding, asymptotically, to $\b \ra
-\infty$,  $\phi \ra -\infty$), then the initial state $\Psi_{in}$ must
represent a configuration which is expanding and with growing  
dilaton, i.e. $\Psi_{in} \sim \psi^+(\b) \psi^-({\fb})$. The quantum
evolution of the initial pre-big bang state is thus represented in this
minisuperspace as the scattering, induced by the effective WDW
potential, of an incoming wave travelling from $-\infty$ along the
positive direction of the axes $\b$ and $\fb$. 

It follows that, in general, there are four  different types of
evolution, depending  on whether the asymptotic outgoing state
$\Psi_{out}$ is a superposition of waves with the same $\pbe$ and
opposite $\pfb$, or with the same $\pfb$ and opposite $\pbe$, and also
depending on the identification of the time-like coordinate in
minisuperspace \cite{7a}.  These four possibilities are illustrated in Fig.
1, where  cases $(a)$ and $(b)$ correspond to $\Psi_{out}^\pm \sim
\psi^+(\b) \psi^\pm({\fb})$, while cases  $(c)$ and $(d)$
correspond to $\Psi_{out}^\pm \sim \psi^-({\fb}) \psi^\pm({\b})$. 

\begin{figure}[t]
\begin{center}
\mbox{\epsfig{file=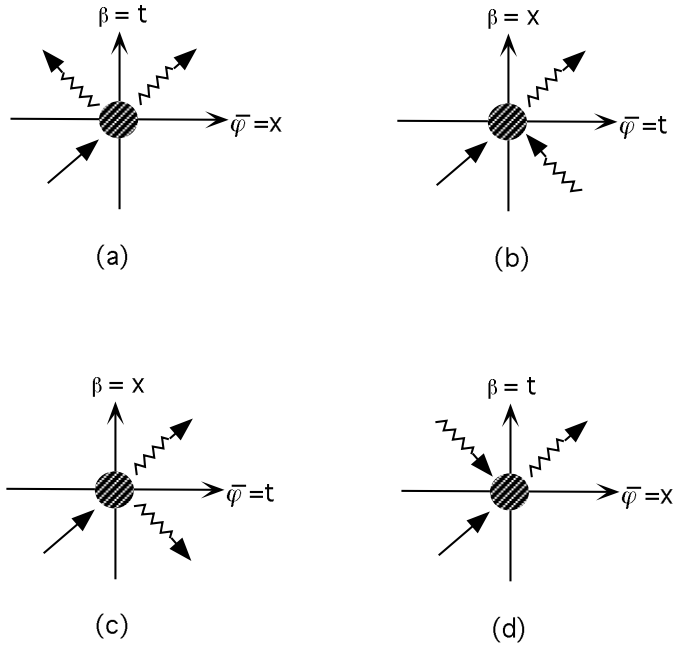,width=90mm}}
\vskip 5mm
\caption{\sl Four different classes of scattering for the initial string
perturbative vacuum (solid line). The two spatial reflections $(a)$ and
$(c)$ describe the transition from an expanding pre-big bang
configuration to an expanding post-big bang and contracting pre-big
bang configuration, respectively. The two Bogoliubov processes $(b)$
and $(d)$ represent the production of universe-antiuniverse pairs from 
the vacuum. In case $(b)$ one universe is expanding, the other
contracting, but they both fall inside the pre-big bang singularity. In
case $(d)$ both universes are expanding, but only one falls inside the
singularity, while the other one survives in the post-big bang regime.} 
\end{center} \end{figure}

The two cases $(a)$ and $(c)$ represent scattering and reflection along
the spacelike axes $\fb$ and $\b$, respectively. In case $(a)$ the
evolution along $\b$ is monotonic, so that the Universe always keeps 
expanding. The incident wave is partially transmitted towards the
pre-big bang singularity (unbounded growth of the curvature and of
the dilaton, $\b \ra +\infty$, $\fb \ra +\infty$), and partially reflected
back towards the low-energy, expanding, post-big bang regime  ($\b
\ra +\infty$, $\fb \ra -\infty$). In case $(c)$ the evolution is monotonic
along the time axis $\fb$, but not along $\b$. So, the incident wave is
totally transmitted towards the singularity ($\fb \ra +\infty$), but in
part as an expanding and in part as a contracting configuration.

In the language of third quantization \cite{11} (i.e., second quantization
of the WDW wave function in superspace) we can say that in case $(a)$
we have the production of expanding post-big bang states 
from the string perturbative vacuum; in case $(c)$, instead,  we have 
the production of contracting pre-big bang states. In both
cases, however, such a production is exponentially suppressed, and the
suppression is proportional to the proper spatial volume of the portion
of Universe emerging from the string perturbative vacuum \cite{7}. 

The other two cases, $(b)$ and $(d)$, are qualitatively different, as the
final state is a superposition of positive and negative energy
eigenstates, i.e. of modes of positive and negative frequency 
with respect to time axes chosen in minisuperspace. In a third
quantization context they represent a ``Bogoliubov mixing", describing
the production of pairs of universes from the vacuum. The mode moving
backwards in time has to be ``re-interpreted", like in quantum field
theory,  as an ``antiuniverse" of positive energy and opposite
momentum (in superspace). Since the inversion of momentum, in
superspace,  corresponds to a reflection of $\dot \b$, the
re-interpretation principle in this context changes expansion into
contraction, and vice-versa. 

Case $(b)$, in particular, describes the production of
universe-antiuniverse pairs -- one expanding, the other contractiong
-- from the string perturbative vacuum. The pairs evolve towards the
strong coupling regime $\fb \ra +\infty$, so both the members of the
pair fall inside the pre-big bang singularity. Case $(d)$ is more
interesting, in our context, since in that case the universe-antiuniverse
of the pair are both expanding: one falls inside the pre-big bang
singularity, the other expands towards the low-energy, post-big bang
regime, and may expand to infinity, representing the birth of a
Universe like ours in a standard Friedman-like configuration. 

Case $(b)$ was discussed in a previous paper \cite{12}: with a simple,
duality-invariant model of potential, it was shown to represent an
efficient  conversion of expanding  into contracting 
internal dimensions, associated to a parametric amplification of the
wave function of the pre-big bang state. In this paper we shall
concentrate on the process illustrated in case $(d)$, already  
conjectured \cite{8} to represent a promising candidate for an efficient
transition from the pre- to the post-big bang regime, but never
discussed in previous papers. To confirm this conjecture, we will provide
here an explicit example in which the production of pairs of universes
containing an expanding post-big bang configuration  may be 
associated to a parametric amplification of the WDW wave function. 

To this purpose we should note, first of all, that for a duality-invariant
dilaton potential the string cosmology Hamiltonian associated to the
action (\ref{1}) is translationally invariant along the $\b$ axis, $[H,
\pbe]=0$: in this case,  an initial expanding configuration keeps
expanding, and the {\em out} state cannot be a mixture of states with
positive and negative eigenvalues of $\pbe$. In order to implement the
process $(d)$ of Fig. 1 we thus need a non-local, duality-breaking
potential, that contains both the metric and the dilaton, but {\em not} in
the combination $\fb$ of eq. (\ref{3}). 

We shall use, in particular, a two-loop dilaton potential induced by an
effective cosmological constant $\La$, i.e. $V \sim \La \exp (2 \phi)$
(two-loop potentials are known to favour the transition to the
post-big bang regime already at the classical level \cite{5,13}, but only
for appropriate repulsive self-interactions with $\La <0$). We shall
assume, in addition, that such a potential is rapidly damped in the large
radius limit $\b \ra +\infty$, and we shall approximate such a
damping, for simplicity, by the Heaviside step function $V \sim \theta
(-\b)$. With this damping we represent
the effective suppression of the cosmological constant, required for the
transition to a realistic post-big bang configuration, and induced by
some physical mechanism that does not need to be specified explicitly, 
for the purpose of this paper. Also, the choice of the cut-off function
$\theta (-\b)$ is not crucial, in our context, and other, smoother
functions would be equally appropriate.

With these assumptions, the WDW equation (\ref{5}) reduces to 
\beq
\left [ \partial^2_{\fb} - 
\partial^2_{ \beta}
+\lambda_s^2\,\Lambda\,\theta(-\b)\,e^{2\sqrt{d}\b} \right ]\,
\Psi(\b,\fb)= 0 , 
\label{8}
\eeq 
and the general solution can be factorized in terms of the 
eigenstates of the momentum $\pfb$, by setting
\beq
\Psi(\b,\fb)=\Psi_k(\b) e^{ik\fb}, ~~~~~~~~~~
\left [\partial^2_{\b}
+k^2-\lambda_s^2\,\Lambda\,\theta(-\b)\,e^{2\sqrt{d}\b} \right ]\,
\Psi_k(\b)= 0.
\label{9}
\eeq
In the region $\b >0$ the potential is vanishing, so that the general
outgoing solution is a superposition of eigenstates of $\pbe$
corresponding to  the positive and negative frequency
modes $\psi_k^\pm$, as in case $(d)$ of Fig. 1. In the opposite region
$\b <0$ the general solution is a combination of  Bessel functions
$J_\nu(z)$, of imaginary index $\nu= \pm ik/ \sqrt{d}$ and argument 
$z= i\la_s \sqrt{\La/d}~e^{\sqrt{d}\b}$. 

We now fix the boundary conditions by imposing that the Universe
starts expanding from the string perturbative vacuum: for $\b
\ra -\infty$, the solution must then reduce to a plane wave
representing a classical, low-energy pre-big bang solution, with
$\pbe=-\pfb =k>0$. In particular, if we use the differential
representation  $\Pi=i\nabla$ for both $\b$ and $\fb$: 
\beq
\Psi_{in}= \lim_{\b \ra -\infty} \Psi(\b,\fb)\sim e^{ik(\fb - \b)} .
\label{10}
\eeq
This choice uniquely determines the WDW wave function as:
\bea
\Psi(\b,\fb)=N_kJ_{-{ik\over\sqrt{d}}}\left(i\la_s
\sqrt{\La\over d}~e^{\sqrt{d}\b}\right) \,e^{ ik\fb}\,,
~~~~~~~~~~~~~~\b&<&0 , \nonumber\\
=\left[A_+(k) e^{ -ik\b}+A_-(k) e^{ik\b}\right]\,e^{ ik\fb}\,,
~~~~~~~~~~~~~~\b&>&0 .
\label{11}
\eea
With the matching conditions at $\b=0$ we
can then compute the Bogoliubov coefficients $|c_\pm(k)|^2 = 
|A_\pm(k)|^2/ |N_k|^2$ determining, in the third quantization
formalism, the number $n_k$ of universes produced from the vacuum, 
for each mode $k$ (here $k$ represents a given configuration in the 
space of the initial parameters). 

In contrast to the tunnelling process discussed in preivious papers
\cite{7,7a}, this process may represent an efficient mechanism of
vacuum decay  since the wave function is parametrically amplified (i.e.,
$n_k\gg 1$) for all $k < \la_s \sqrt\La$. To illustrate this point we have
numerically integrated eq. (\ref{9}), and plotted in Fig. 2 the evolution
in superspace of the real part of the wave function, for different
configurations of initial momentum $k$ (the behaviour of the imaginary
part is qualitatively similar). 

\begin{figure}[t]
\begin{center}
\mbox{\epsfig{file=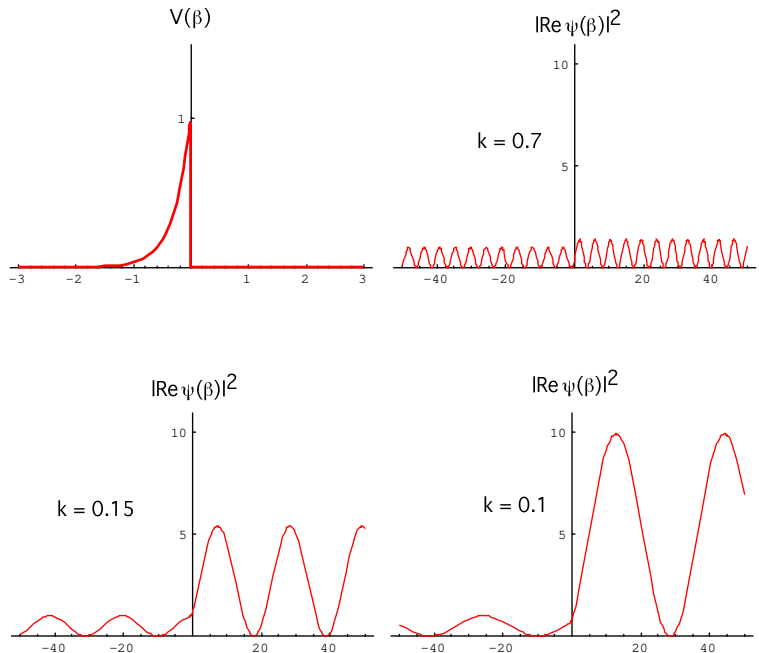,width=100mm}}
\vskip 5mm
\caption{\sl The first plot represents the effective potential of the
WDW equation (\ref{9}), for $d=3$, in units of $\la_s^2 \La$. The other
plots represent the evolution in superspace of 
$\left|\Re{\rm e}\Psi_k(\b)\right|^2$,  obtained by solving eq. (\ref{9})
with the initial boundary condition (\ref{10}), for different values
of  $k$. We have used for all modes the same normalization, $|\Psi_k|^2
=1$ at $\b \ra -\infty$, to emphasize that the amplification is more
effective at lower frequencies.} 
\end{center}
\end{figure}

It may be interesting to note that the amplification is smaller at higher 
frequencies or -- to use the language of cosmological  
perturbation theory -- the pairs of universes are produced with
a decreasing spectrum .  This result has a quite reasonable 
interpretation, once we express the momentum $k$ in terms of the
physical parameters of the final geometric configuration. Indeed, from
the definitons (\ref{3}) and (\ref{4}) we find, for a realistic
transition occurring at the string scale, $\dot \b \sim \la_s$, that $k
\sim (\Om_3/\la_s^3)g_s^{-2}$, where $\Om_3 = a^3 \int d^3x$ is the
proper spatial volume emerging from the transition in the post-big
bang regime, and $g_s =e^\phi/2$ is the string coupling, parametrized
by the dilaton. The condition of parametric amplification, 
\beq
k \sim \left(\Om_3\over \la_s^3\right){1\over g_s^2} \laq \la_s \sqrt
\La, 
\label{12}
\eeq
implies that the transition is strongly favoured for configurations of
small enough spatial volume in string units, large enough coupling
$g_s$, and/or large enough cosmological constant $\La$, in string units
(in agreement with previous results \cite{7,12}). 

For $k \gg \la_s \sqrt{\La}$ the wave function does not  ``hit" the
barrier, and there is no parametric amplification. The inital state runs
almost undisturbed towards the singularity, and only a small,
exponentially suppressed fraction is able to emerge in the post big bang
regime. In the context of third quantization this process can still be
described as the production of pairs of universes, but the number of
pairs is now exponentially damped, $n_k \sim \exp(-k/\la_s \sqrt\La)$,
with a  Boltzmann factor corresponding to a ``thermal bath" of
universes, at the effective temperature $T \sim \sqrt \La$ in
superspace. 

In view of the above results, we may conclude that the decay of the
string perturbative vacuum, if triggered by an appropriate,
duality-breaking dilaton potential, can efficiently proceed via the
parametric amplification of the WDW wave function in superspace, and
can describe the birth of our Universe as a forced production of pairs
from the vacuum fluctuations. One member of the pair disappears into
the pre-big bang singularity, the other bounces back towards the
low-energy region. The resulting effect is a net flux of universes that
may escape to infinity in the post-big bang regime (see Fig. 3). This
effect is similar to the quantum emission of radiation from a black
hole \cite{14}, with the difference that the quanta produced in pairs
from the vacuum are separated not by the black-hole horizon, but by 
the ``Hubble" horizon associated to the ``accelerated" variation of the
dilaton in minisuperspace. 

\begin{figure}[h]
\begin{center}
\mbox{\epsfig{file=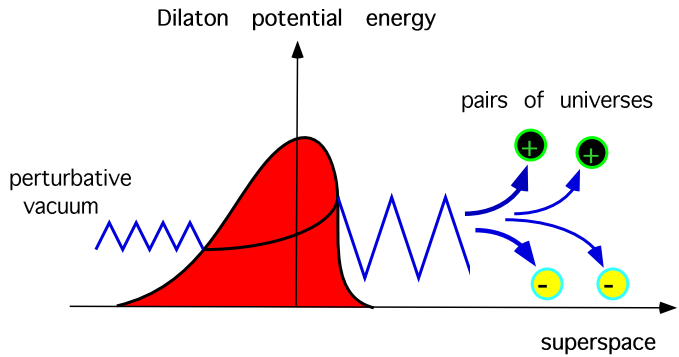,width=90mm}}
\vskip 5mm
\caption{\sl Birth of the universe represented as an anti-tunneling
effect in superspace, or as a process of pair production from the string
perturbative vacuum.}
\end{center}
\end{figure}

\end{document}